\documentclass[sigconf]{acmart}

\usepackage{lineno}
\modulolinenumbers[5]
\usepackage[utf8]{inputenc}
\usepackage[T1]{fontenc}
\usepackage{amsmath,amsfonts,mathrsfs}
\usepackage{algorithmic}
\usepackage{graphicx}
\usepackage{array}
\usepackage[caption=false,font=footnotesize,labelfont=sf,textfont=sf]{subfig}
\usepackage{textcomp}
\usepackage{booktabs} 
\usepackage{multirow}
\usepackage{listings}
\def\BibTeX{{\rm B\kern-.05em{\sc i\kern-.025em b}\kern-.08em
    T\kern-.1667em\lower.7ex\hbox{E}\kern-.125emX}}

\usepackage{xspace}
\newcommand{\ie}{i.e.,\xspace}
\newcommand{\eg}{e.g.,\xspace}

\xspaceaddexceptions{=}

\usepackage[acronym,toc,nonumberlist]{glossaries}
\makenoidxglossaries
\newacronym{fft}{FFT}{Fast Fourier Transform}
\newacronym{spmv}{SPMV}{Sparse matrix-vector multiplication}
\newacronym{pr}{PR}{Page-Rank}
\newacronym{bfs}{BFS}{Breadth-First Search}
\newacronym{hpc}{HPC}{High-Performance Computing}
\newacronym{csr}{CSR}{Control and Status Register}
\newacronym{isa}{ISA}{Instruction Set Architecture}
\newacronym{vl}{VL}{Vector Length}

\usepackage{listings} 
\definecolor{col1}{HTML}{1E88E5}
\definecolor{col2}{HTML}{D81B60}
\definecolor{col3}{HTML}{43A047}
\definecolor{col4}{HTML}{F4511E}
\definecolor{col5}{HTML}{E259FF}

\usepackage{pifont}
\RequirePackage[normalem]{ulem}
\RequirePackage{color}\definecolor{RED}{rgb}{1,0,0}\definecolor{BLUE}{rgb}{0,0,1}

\providecommand{\DIFdel}[1]{{\protect\color{RED}\sout{#1}}}
\providecommand{\DIFdel}[1]{}

\setcopyright{acmcopyright}
\copyrightyear{2023}
\acmYear{2023}
\acmDOI{XXXXXXX.XXXXXXX}

\acmConference[SC23]{The International Conference for High Performance Computing, Networking, Storage, and Analysis}
{November 12--17, 2023}{Denver, CO}

\begin{document}

%
\title[Short reasons for long vectors in HPC CPUs]{Short reasons for long vectors in HPC CPUs:\\a study based on RISC-V}

%
\author{Pablo Vizcaino}
\email{pablo.vizcaino@bsc.es}
\orcid{4444-4444-4444}
\affiliation{%
  \institution{Barcelona Supercomputing Center}
  \streetaddress{C. Jordi Girona, 29}
  \city{Barcelona}
  \country{Spain}
  \postcode{08034}
}
\author{Georgios Ieronymakis}
\email{ieronym@ics.forth.gr}
\orcid{0000-0001-9776-985X}
\affiliation{%
  \institution{FORTH-ICS}
  \streetaddress{N. Plastira 100}
  \city{Heraklion, Crete}
  \country{Greece}
  \postcode{70013}
}

\author{Nikolaos Dimou}
\email{ndimou@ics.forth.gr}
\orcid{0000-0002-5788-0892}
\affiliation{%
  \institution{FORTH-ICS}
  \streetaddress{N. Plastira 100}
  \city{Heraklion, Crete}
  \country{Greece}
  \postcode{70013}
}

\author{Vassilis Papaefstathiou}
\email{papaef@ics.forth.gr}
\orcid{0000-0002-5443-6470}
\affiliation{%
  \institution{FORTH-ICS}
  \streetaddress{N. Plastira 100}
  \city{Heraklion, Crete}
  \country{Greece}
  \postcode{70013}
}
\author{Jesus Labarta}
\email{jesus.labarta@bsc.es}
\orcid{4444-4444-4444}
\affiliation{%
  \institution{Barcelona Supercomputing Center}
  \streetaddress{C. Jordi Girona, 29}
  \city{Barcelona}
  \country{Spain}
  \postcode{08034}
}

\author{Filippo Mantovani}
\email{filippo.mantovani@bsc.es}
\orcid{1234-5678-9012}
\affiliation{%
  \institution{Barcelona Supercomputing Center}
  \streetaddress{C. Jordi Girona, 29}
  \city{Barcelona}
  \country{Spain}
  \postcode{08034}
}

%
\renewcommand{\shortauthors}{P. Vizcaino, et al.}

%
\begin{abstract}
%
%
%

For years, SIMD/vector units have enhanced the capabilities of modern CPUs in High-Performance Computing (HPC) and mobile technology. Typical commercially-available SIMD units process up to 8 double-precision elements with one instruction. The optimal vector width and its impact on CPU throughput due to memory latency and bandwidth remain challenging research areas.
%
%
This study examines the behavior of four computational kernels on a RISC-V core connected to a customizable vector unit, capable of operating up to 256 double precision elements per instruction.
The four codes have been purposefully selected to represent non-dense workloads: SpMV, BFS, PageRank, FFT.
The experimental setup allows us to measure their performance while varying the vector length, the memory latency, and bandwidth.
%
%
Our results not only show that larger vector lengths allow for better tolerance of limitations in the memory subsystem but also offer hope to code developers beyond dense linear algebra.
%
%


\end{abstract}


\maketitle

\section{Introduction and related work}\label{secIntro}

Various approaches to vector computing exist, with the prevalent ``classical'' Single Instruction, Multiple Data (SIMD) method employed by x86 architectures. Recently, two pivotal technological developments have driven the progression of research and software advancement within the realm of vector computing: the Arm Scalable Vector Extension (SVE) and the NEC Aurora SX architecture. Arm's innovation introduces a more adaptable and sophisticated approach to vectorizing code, supporting vector lengths of up to 32 double precision elements, with a register size of 2 kbits, as detailed by Rico et al. in~\cite{rico_arm_2017}. On the other hand, the NEC Aurora SX takes an even more radical design stance, presenting a CPU equipped with vector registers accommodating a staggering 256 double precision elements, spanning 16 kbits, as elaborated by Takahashi et al. in~\cite{takahashi_performance_2023}.

These advancements raise significant inquiries concerning vector architectures, as explored by Ishii et al. in~\cite{ishii_architectural_2013}. Key among these are inquiries into the optimal vector length, the ideal vector register size, and the most favorable memory performance metrics, encompassing both bandwidth and latency, that should complement a vector architecture. Beyond the purely architectural questions, the intricacies of simulating and benchmarking vector architectures also come to the fore, as expounded upon by Ramirez et al. in~\cite{ramirez_risc-v_2020}.

Indeed, diverse scientific communities are now intently investigating the potential of long-vector computing as a means to speedup their computational endeavors. Importantly, these pursuits often extend beyond the confines of classical high-performance computing (HPC) operations, as demonstrated \eg by Valassi et al. in~\cite{valassi_design_2021} and by Diehl et al. in~\cite{diehl_simulating_2023}. 
In our study, we embark on the task of quantifying the precise influence of vector length, memory bandwidth, and memory latency on distinct code implementations. Our objective is to substantiate the claim that vector architectures, particularly those leveraging extended vectors, offer robust solutions for gaining performance while tolerating memory latency and memory bandwidth.

Additionally, since our analysis delves into kernels characterized by non-dense computations, we endeavor to convey a broader message: the realm of scientific software encompasses domains beyond those rooted in dense linear algebra and vector architectures offer hope to all scientific application developers.

Lee et al., in~\cite{lee_test-driving_2023} present an overview of the RISC-V technologies leveraging the RISC-V Vector Extension available from the industry and the academia.
For the evaluation part of our work, we have chosen to focus on the hardware development of a RISC-V-based design that targets the HPC domain and is developed within the European Processor Initiative project.
The system leverages a RISC-V core coupled with a Vector Processing Unit implementing the RISC-V Vector Extension RVV-v0.7.1 and it allows to configure \Gls{vl} and memory parameters.

The main contributions of this paper are:
{\em i)} we present a flexible FPGA-based setup (Software Development Vehicle) including a RISC-V core connected to a configurable VPU on which we can co-design hardware and software in a flexible and fast way;
{\em ii)} we quantify the effects of \Gls{vl}, memory latency, and memory bandwidth on a real RISC-V CPU implemented on FPGA;
{\em iii)} we study four different kernels that are not-dense: SpMV, BFS, PR and FFT.

The rest of this paper is structured as follows:
Section~\ref{secExperimentalSetup} presents the experimental setup on which we performed the measurements presented in this paper.
Section~\ref{secMethodology} explains the methodology adopted, the software setup and the codes selected for the benchmarking.
In Section~\ref{secEvaluation} we provide details about the evaluation, reporting performance measurements when varying \Gls{vl}, memory latency and memory bandwidth.
We conclude the paper with our findings and comments in Section~\ref{secConclusions}.

\section{Experimental Setup}\label{secExperimentalSetup}



In this section we describe our experimental setup called FPGA Software Development Vehicle (FPGA-SDV), which is developed within the European Processor Initiate (EPI) project%
\footnote{European Processor Initiative. https://www.european-processor-initiative.eu/} 
and emulates parts of the EPI European Accelerator (EPAC) design.
The parts of EPAC that we emulate on the FPGA design include:
\begin{itemize}
  \item Atrevido, a RISC-V super-scalar core developed by SemiDynamics\footnote{SemiDynamics. \url{https://semidynamics.com/}}.
  \item Vitruvius Vector Processing Unit (VPU)~\cite{minervini2023vitruvius} developed by BSC\footnote{Barcelona Supercomputing Center. \url{https://bsc.es/}},  
        which is tightly coupled to the core, contains eight lanes and is capable of operating on vectors of 256 double-precision elements (\ie 16384-bit wide vector registers). 
        Each lane includes a Floating Point Unit (FPU) developed by the University of Zagreb~\cite{kovavc2023faust}.
  \item Network on Chip (NoC) with 2D Mesh topology developed by EXTOLL\footnote{EXTOLL. \url{http://extoll.de/}}.
  \item Shared L2 Cache developed by FORTH-ICS\footnote{FORTH-ICS CARV. \url{https://www.ics.forth.gr/carv}} which is tightly coupled with a MESI-based coherence Home Node developed by Chalmers\footnote{Chalmers University. \url{https://www.chalmers.se/en/departments/cse/}}, collectively called L2HN.
\end{itemize}

\begin{figure}[htbp]
  \centering
  \includegraphics[width=.55\linewidth]{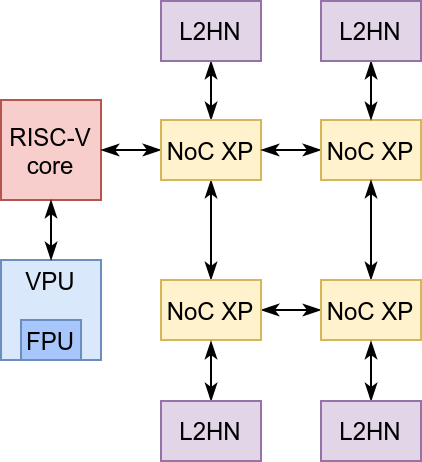}
  \caption{Simplified architecture of the FPGA-SDV}
  \label{figVecSdvBlockDiagram}
\end{figure}

Figure~\ref{figVecSdvBlockDiagram} provides a simplified block diagram of each components of FPGA-SDV: Atrevido in red, VPU+FPU in blue, NoC in yellow and L2HN in purple.

\begin{figure}[htbp]
  \centering
  \includegraphics[width=.95\linewidth]{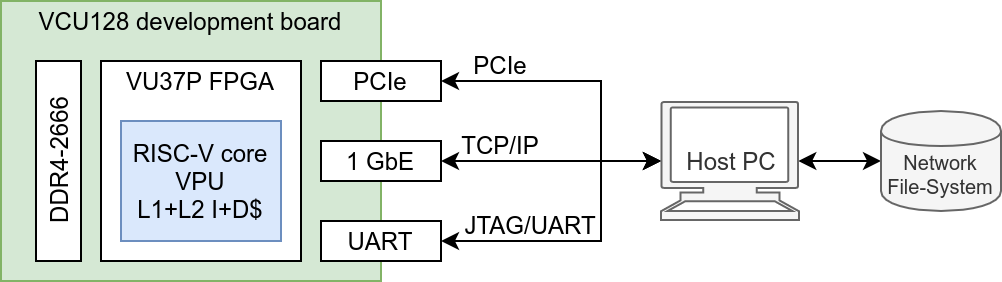}
  \caption{Connection between host server and VCU128 board: schematic view}
  \label{figSdvComponents}
\end{figure}

Our setup comprises of a Virtex UltraScale+ HBM VCU128 PCIe FPGA board\footnote{Xilinx VCU128. \url{https://www.xilinx.com/products/boards-and-kits/vcu128.html}} 
and a AMD Ryzen5 5600 host server with 32~GB of DDR4-3200 memory, both mounted on a Mini-ITX motherboard. The server runs Ubuntu Server Linux 20.04 with local storage and a mounted Network Filesystem.
The VCU128 board includes a VU37P FPGA device 
with 8~GB of integrated HBM memory and 4.5~GB of external DDR4 memory.
Figure~\ref{figSdvComponents} shows the connectivity between the host server and the VCU128 board.

In this work we emulate on the FPGA one instance of Atrevido with VPU and four instances of L2HN connected via a $2\times2$ Mesh NoC.
The host server communicates with the VCU128 primarily through PCIe Gen3 x16, 1~Gbit Ethernet, and other low-speed interfaces like JTAG and UART. 
A typical scenario for experimenting with the FPGA-SDV includes the following steps:

\begin{itemize}
  \item Use the host to program the VCU128 bitstream with the emulated system via JTAG.
  \item Use the host PCIe interface to load Linux images into the VCU128 DDR4 memory, perform any additional register configurations 
        like modifying the memory latency as described in Section~\ref{sec:LatCtrl}, and initiate boot on the emulated system.
  \item Access the emulated system's Linux terminal via UART or SSH via Ethernet.
  \item Mount a Network Filesystem (NFS) via Ethernet in order to share files between the host and the emulated system, run benchmarks, and perform tests.
\end{itemize}

\subsection{Variable Vector Length}



The RISC-V Vector Extension\footnote{RISC-V "V" Extension. \url{https://github.com/riscv/riscv-v-spec/blob/master/v-spec.adoc}} 
has some characteristics that differentiate it from other vector extensions, such as Intel's AVX or ARM Neon.
The \Gls{vl} is not constrained by the \Gls{isa}; it is implementation-specific like ARM's SVE extension.
Additionally, the \Gls{vl} can change on runtime.
A code can execute some instructions with a given \Gls{vl} and some other instructions with a shorter or longer \Gls{vl}, which means that the architecture adapts to the codes seamlessly without predicates or loop prologues/epilogues.
This also allows the compiler to generate \Gls{vl}-agnostic code, meaning that the binaries do not need to know the machine's maximum implemented \Gls{vl}, thus enabling portability.

Inside the core we can find a custom \Gls{csr} that contains the maximum \Gls{vl} of the machine.
Usually this value would be hard-coded in the hardware, but having it in a \Gls{csr} lets us programmatically modify it with lower values to run interesting experiments.
Lowering the \Gls{vl} may degrade performance, but allows us to quantify its effect and see how it interacts with other parameters of the core.

\subsection{Latency Controller} \label{sec:LatCtrl}



The Latency Controller is a hardware module which can artificially introduce additional latency to the DDR4 memory subsystem. 
In our experimental setup the FPGA emulated Atrevido + VPU + NoC + $2 \times 2$ L2HN system runs at a lower frequency compared to DDR4 of the VCU128 FPGA, \ie 50~MHz compared to 333~MHz. 
Therefore, from the emulated system's perspective the memory requests appear to complete with lower latency compared to a typical production system; 
the minimum memory access latency observed in the FPGA emulated system at 50 MHz is approximately 50 clock cycles. 
Thus, this module primarily serves two goals: 
{\em (a)} to emulate the memory latency of a typical system and 
{\em (b)} evaluate how latency affects the behavior and performance of benchmarks running on the emulated system.
Practically, the Latency Controller offers two functionalities: 
{\em (a)} stalls reads and writes to DDR4 memory for a user-defined number of clock cycles in a pipelined fashion, 
{\em (b)} provides a software configurable interface so that users can modify the memory latency in a programmable and dynamic way 
on the emulated system without re-configuring the FPGA.

\subsection{Bandwidth Limiter} \label{sec:BwLim}



The Bandwidth Limiter is a hardware module that can throttle the bandwidth of the DDR4 memory subsystem. 
More specifically, this module operates in time windows and permits only a limited number of memory requests per time window. 
For example, to throttle the bandwidth at 33\% of the peak only two registers need to be set: the value of the numerator to 1 and the value of the denominator to 3 to achieve the fraction 1/3. 
Then, the module allows only 1 memory request to be issued per time window, which is 3 clock cycles.
The Bandwidth Limiter operates similarly to the Latency Controller and provides the same software configurable interface for flexibility,
so that users can experiment with different bandwidth rates quickly without re-configuring the FPGA.

\section{Methodology}\label{secMethodology}


The core inside the FPGA runs a regular Ubuntu Server 20.04 with local storage and a mounted Network Filesystem.
We compile our codes using an LLVM-based compiler developed at the BSC\footnote{\url{https://repo.hca.bsc.es/gitlab/rferrer/llvm-epi}}, which can vectorize for RISC-V using either automatic vectorization or manual vectorization via intrinsics or builtins.
The codes we evaluate have been vectorized using a combination of both methods; we use automatic vectorization for short or simple regions of code and intrinsics to perform fine-grained optimizations.

\subsection{Codes Selected for the Evaluation}

We employ four computationally relevant codes in our evaluation.
All of them are vectorized for RISC-V using the LLVM-based compiler with vectorization support.

First, we use a vectorized \Gls{spmv} targeting long-vector architectures~\cite{gomez_crespo_optimizing_2020}.
The C sources can be accessed upon request in BSC repositories\footnote{\url{https://repo.hca.bsc.es/gitlab/cgomez/spmv-long-vector}}.
The input used for this evaluation is the ``CAGE10'' matrix\footnote{\url{https://sparse.tamu.edu/vanHeukelum/cage10}}.
\Gls{spmv} is a relevant code in \Gls{hpc} because it behaves more similarly to real scientific applications than artificial benchmarks due to its use of sparse data structures and being memory bound.
It is normally used to highlight the efficiency of the memory subsystem of an architecture.

Secondly, we use long-vector implementations of two graph algorithms, \Gls{bfs} and \Gls{pr}~\cite{vizcaino_serrano_implementing_2023}.
The C++ sources can also be accessed upon request\footnote{\url{https://repo.hca.bsc.es/gitlab/pvizcaino/graph-v}}.
Graph algorithms are a relevant \Gls{hpc} topic in various fields like networks, infrastructure, social circles, or internet webpages.
\Gls{bfs} is one of the best-known examples and a building block of many other algorithms.
\Gls{pr} presents slightly more computational intensity, and it is used to assign scores to nodes in a graph.
Web browsers such as Google use it to determine a webpage's relevance.
Both codes used a graph of $2^{15}$ nodes for the evaluation.

Finally, we use a vectorized implementation of the \gls{fft} kernel~\cite{fftp}.
The C sources can be accessed upon request\footnote{\url{https://repo.hca.bsc.es/gitlab/pvizcaino/fftv}}.
For the evaluation, we selected an FFT size of 2048 elements.
The \Gls{fft} is a relevant code because it presents both arithmetic intensity and complex memory access patterns, which makes it a challenge for vector architectures.
It is also used as a building block in many scientific codes including signal or image processing, numerical analysis, genomics, or astronomy.

\subsection{Tools Used for Measurements}

To obtain fine-grained measurements and mitigate OS noise, we read the hardware counter that counts CPU cycles. We then calculate the average of these measurements over five runs. As the variation among the runs is below 3\%, we have chosen to exclude error bars to prevent unnecessary cluttering of the plots.


\section{Evaluation}\label{secEvaluation}

In this section, we evaluate the behavior of the four codes when we artificially introduce latency, throttle bandwidth, or alter the maximum \Gls{vl} of our system.
The analysis of vector architectures often revolves around the benefit of operating many data elements simultaneously, without giving much regard to their interaction with the memory.
Using long vectors, we can pack hundreds of memory requests in a single instruction, thus dramatically reducing the number of times we pay the latency to the memory, while also highly utilizing the bandwidth since each transaction is moving many more data elements compared to scalar accesses.

\subsection{Adding latency}\label{subsecLatency}

Using the tools described in Section~\ref{sec:LatCtrl}, we can add an arbitrary number of cycles to the latency between the L2 cache and the main memory.
This way, we can simulate a system with more congestion (\eg with more cores) or with a longer path to the memory and study its effects on scalar and vectorized applications.

With these tests, we aim to showcase that long-vector architectures are more resistant to memory latency than traditional scalar machines.
For each code, we study both the scalar and vector implementations. For the vector ones, we also modify the maximum \Gls{vl} of the system in order to study its relation with the latency.

\begin{figure}[!htbp]
  \centering
  \includegraphics[width=1.0\linewidth]{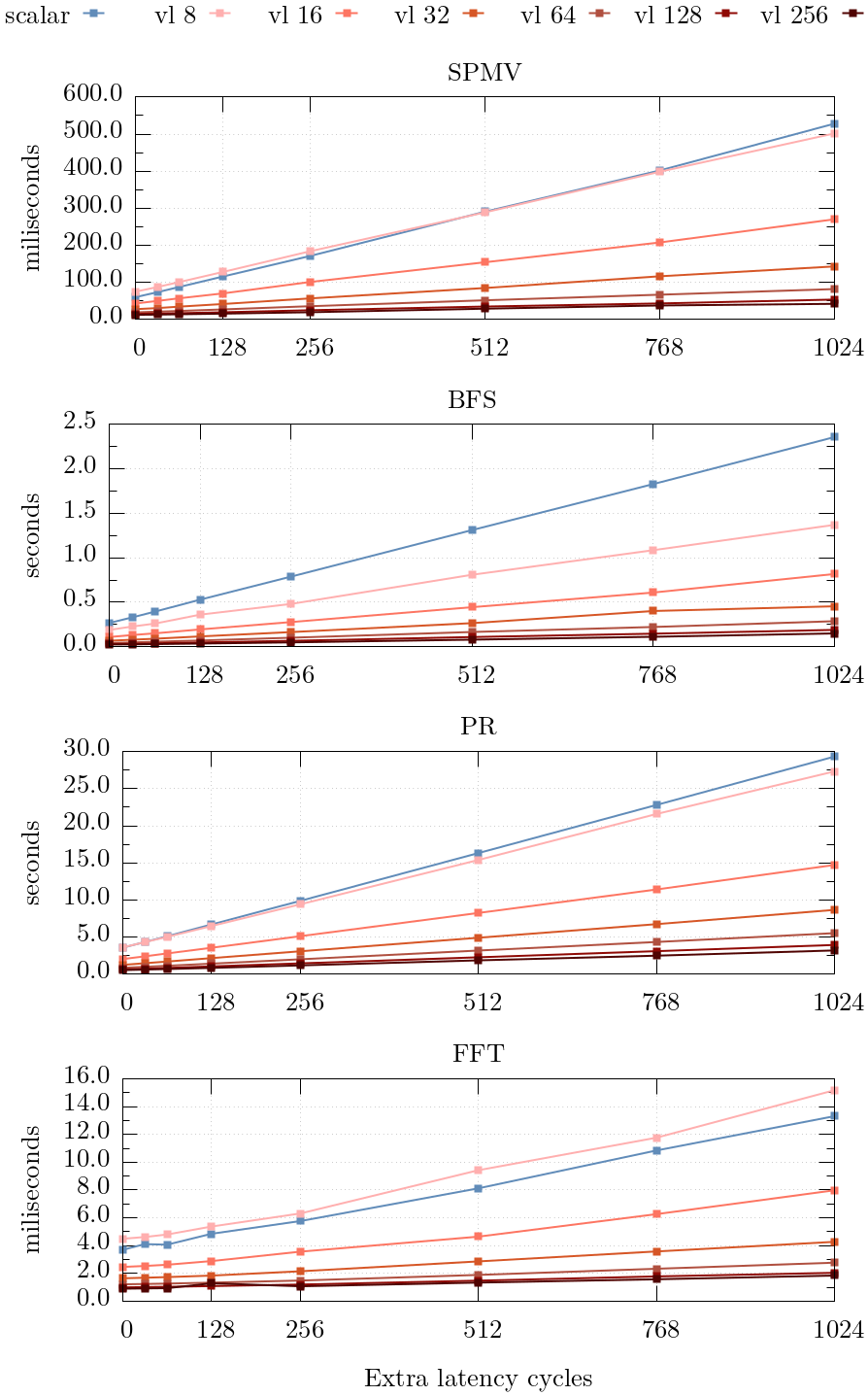}
  \caption{Execution time of the four evaluated codes depending on the added latency, with blue series for scalar implementations and a red gradient series for vector implementations with increasing \Gls{vl}}
  \label{figLatencyVlTime}
\end{figure}

Figure~\ref{figLatencyVlTime} contains four plots with the measured execution time for each of the codes on the $y$-axis.
The $x$-axis report the extra latency cycles that have been added by the Latency Controller described in Section~\ref{sec:LatCtrl}.
Each plot shows the scalar implementation in blue, and the vector runs with increasing \Gls{vl} in a red gradient: the larger the \Gls{vl} the darker the red. 
The values of \Gls{vl} considered are expressed in units of double-precision elements: \ie \Gls{vl}=8 means vector register width of 512~bits (light red) while \Gls{vl}=256 means vector register width of 16~kbits (dark red).
This color pattern applied to the \Gls{vl} is maintained for the rest of the document for an easier comparison of the plots.


As anticipated, all the plots in Figure~\ref{figLatencyVlTime} demonstrate an increase in execution time with the addition of extra latency. 
However, Figure~\ref{figLatencyVlTime} also distinctly illustrates the substantial advantage that the vector code gains from a higher \Gls{vl}. 
When examining the slopes of the data series, it is evident that as we decrease the \Gls{vl}, the slopes become steeper. 
This implies that the scalar and low \Gls{vl} implementations are more adversely impacted (\ie perform worse) by the introduced latency, causing their performance to degrade more rapidly compared to executions with larger \Gls{vl} values.

\begin{figure}[!htbp]
  \centering
  \includegraphics[width=1.0\linewidth]{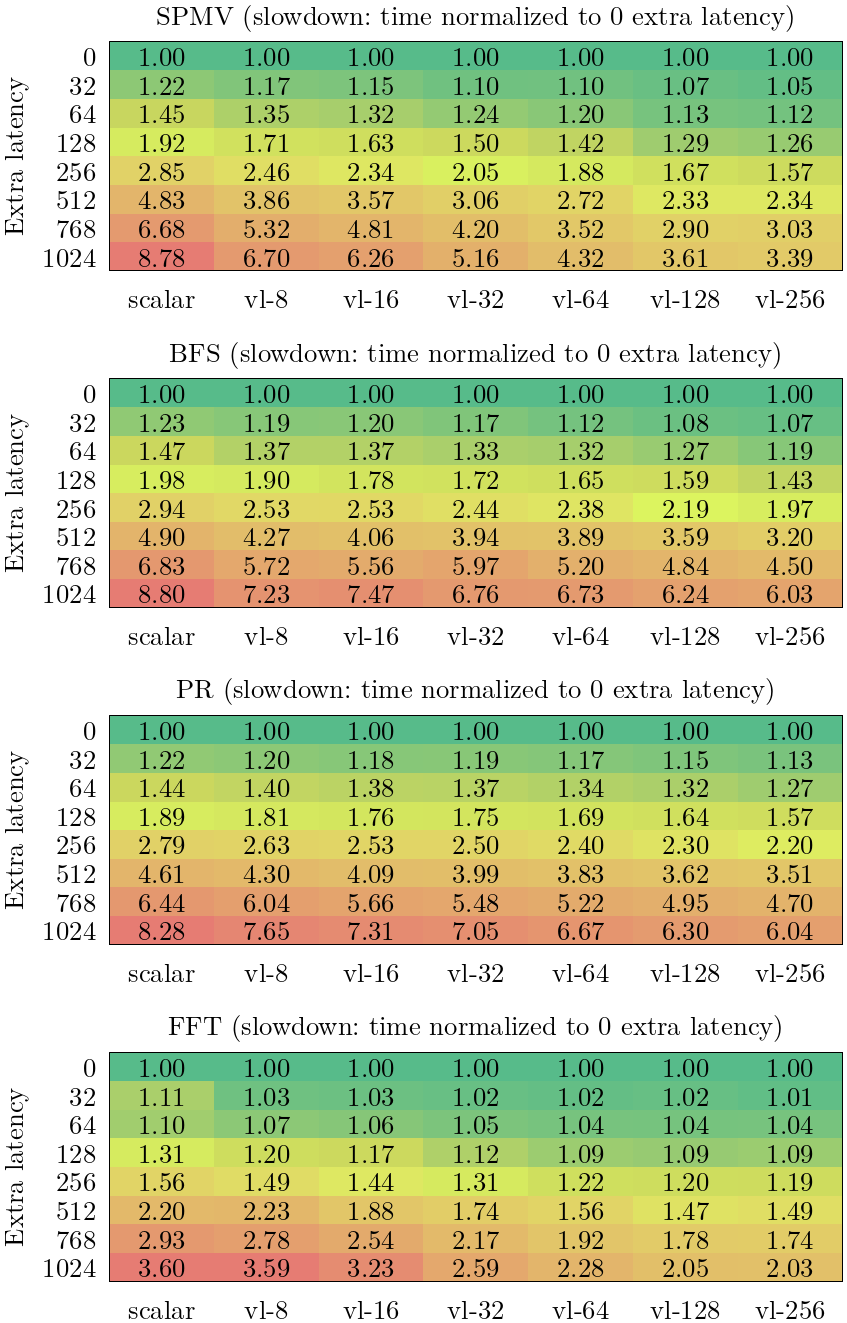}
  \caption{Execution time of the four evaluated codes normalized to their run with 0 extra latency for each implementation. The color gradient goes from minimum slowdown (green) to maximum slowdown (red) for each table}
  \label{figLatencyVlSlowdown}
\end{figure}

We propose an alternative data visualization to better study the effect of the \Gls{vl} on the latency resistance in Figure~\ref{figLatencyVlSlowdown}.
We show one table per each code, with the different implementations (scalar and with different \Gls{vl} values) as columns and the extra latency cycles as rows.
The values in each cell of the tables are the execution times normalized against each implementation's run with no additional latency.
This way, for each implementation, we show how many times slower they run depending on the extra latency added.
In each table we color-code this slowdown from green for the best case (lowest slowdown) to red for the worst case (highest slowdown).


Looking at the tables row by row, we can see that as we add latency, the tables turn red as the slowdown increases.
The key observation is that for any added latency row, the slowdown diminishes when we increase the \Gls{vl}, with the minimum slowdown at the right-most column.

Using the \Gls{spmv} as an example, adding 32 cycles of latency the scalar code runs $1.22\times$ slower, while the vector implementation with $vl=256$ only runs $1.05\times$ slower. 
This is even more pronounced when adding 1024 cycles of latency, with a slowdown of $8.78\times$ compared to $3.39\times$.

\subsection{Limiting bandwidth}\label{subsecBandwidth}

The second use of the tools from  Section~\ref{sec:BwLim} is limiting the bandwidth, from 64 Bytes per cycle downwards.
This lets us study how vectors use the bandwidth provided by the system.

A scalar architecture normally needs many CPUs sending requests to the memory to fully saturate the bandwidth that the system can provide.
On the other hand, long vectors can reach this saturation quicker using only one core, making a more efficient use of the system with a lower number of cores.

\begin{figure}[!htbp]
  \centering
  \includegraphics[width=1.0\linewidth]{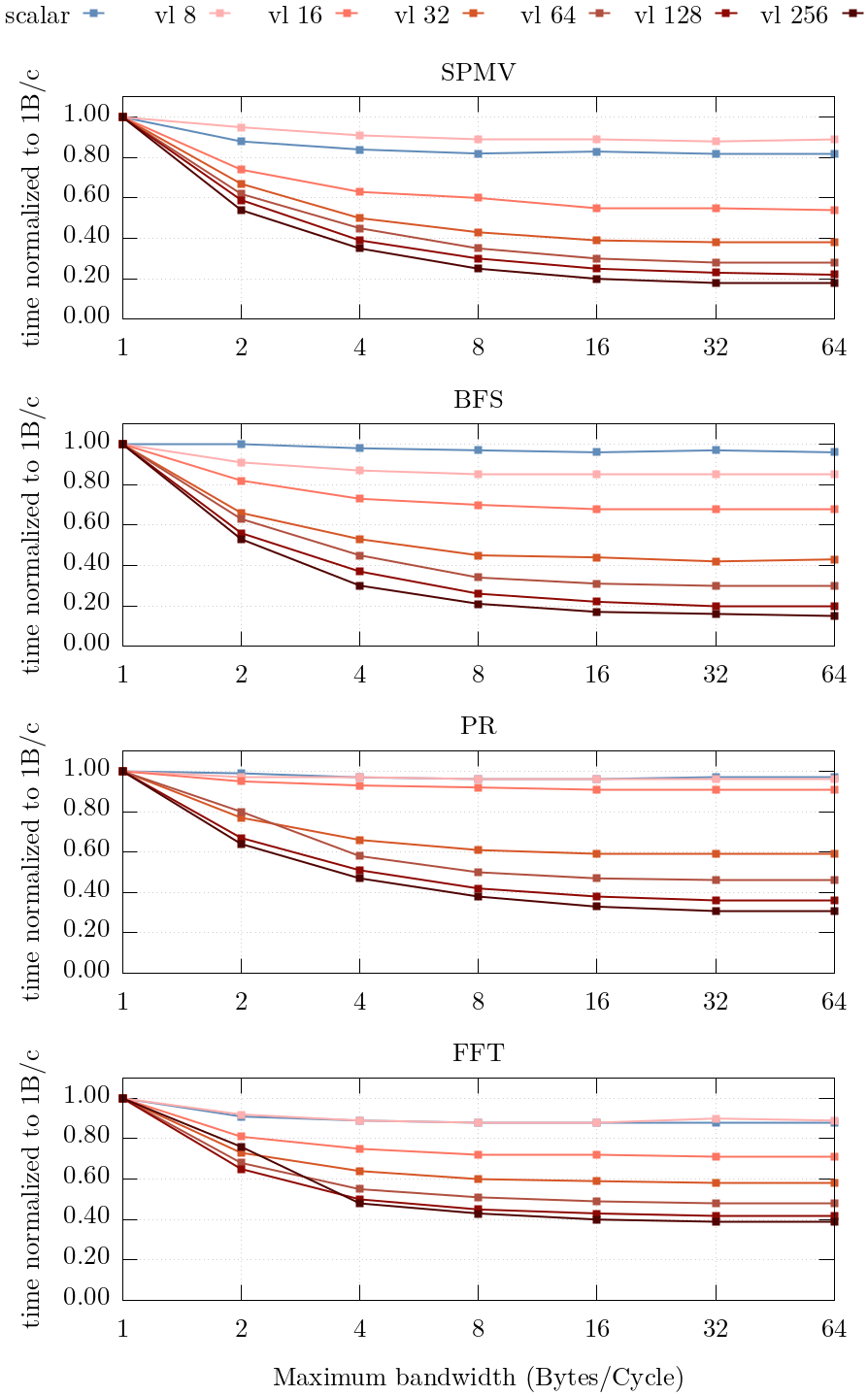}
  \caption{Execution time of the four codes depending on the limited bandwidth, normalized to the run with a limit of $1$ Byte/Cycle per each implementation}
  \label{figBandwidthVlTime}
\end{figure}

Figure~\ref{figBandwidthVlTime} shows the same four codes as before, once again with a blue series for the scalar implementation and a red gradient of series for the vector implementations with increasing \Gls{vl} values.
The $x$-axis represents the maximum bandwidth we limit with our Bandwidth Limiter tool. The values of the $x$-axis are expressed in Bytes per cycles, ranging from 1 to 64 Bytes per cycles in steps of power of 2.
The $y$-axis is the execution time normalized against the same run with a maximum bandwidth of 1 Byte per cycle.
This means that lower values on the $y$-axis are preferred since they represent a shorter execution time.

The plots on Figure~\ref{figBandwidthVlTime} showcase that the scalar versions do not take advantage of bandwidths higher than $1$ or $2$ Bytes/Cycle.
%
This is seen as the time curve from the scalar implementations (blue) flattening and creating a plateau at those values.
The vector code with $vl=8$ behaves similarly to the scalar code, taking little advantage of bandwidths higher than $2$ or $4$ Bytes/Cycles
On the other hand, as we increase the maximum \Gls{vl} of the machine (darker series in the plots), the plateau of the execution time is reached with higher bandwidth values.
This implies that codes with larger \Gls{vl} benefit more from having a system with higher bandwidth.
In other words, Figure~\ref{figBandwidthVlTime} shows that a single core with a vector unit supporting larger \Gls{vl} takes more benefits from being connected to a memory subsystem with higher bandwidth than a scalar core (or a core with shorter \Gls{vl}).

\section{Conclusions}\label{secConclusions}


In this paper we presented an FPGA setup, called FPGA-SDV implementing a scalar RISC-V core coupled with a configurable vector unit with 8 lanes able to process vector of up to 256 double precision elements (16~kbits).
The setup is configured so that a user can configure the maximum vector length (applying values from 8 up to 256 double-precision elements), the memory bandwidth (with values from 1 to 64 Bytes per cycle), and the memory latency (adding an arbitrary number of extra-cycles of latency for accessing the main memory).

We used this setup to evaluate four kernels that are relevant in \Gls{hpc} but are not strictly related to dense-algebraic workflow: we considered \Gls{spmv}, \Gls{bfs}, \Gls{pr}, and \Gls{fft}.

On those non-dense codes, we showcase two often overlooked benefits of long vector architectures:
{\em i)} tolerate memory latency; 
{\em ii)} taking advantage of higher memory bandwidths with a single core;

We highlight the potential of the FPGA-SDV methodology to prove these benefits.
The flexibility of our system lets us easily tweak architectural parameters on real hardware, such as the maximum vector length, the bandwidth of the system, or the latency to the main memory.
%
%
Coupling these features with the methodology described in our previous paper~\cite{mantovani_software_2023} allow us to study \Gls{hpc}-relevant codes and to implement a co-design cycle that provides valuable insights to scientists developers, compiler/system software developers, and hardware architects.

When we degrade the performance of the codes by adding extra memory latency to our system, we observed that the vectorized implementations are less impaired than the scalar ones. 
This difference is accentuated when the vector implementations use a large vector length, thus showing that long vectors are more resistant to high latencies.
In our experimental setup, we also observed that while the single core scalar implementations do not benefit from a bandwidth higher than $1$ or $2$ Bytes/Cycle, the long vector implementations can naturally use bandwidths of $32$ or $64$ Bytes/Cycle.

\begin{acks}
This research has received funding from the European Commission via the Horizon Europe research and innovation funding programme, under grant agreement 101092993 (RISER) and the European High Performance Computing Joint Undertaking (JU) under Framework Partnership Agreement No 800928 (European Processor Initiative) and Specific Grant Agreement No 101036168 (EPI SGA2). The JU receives support from the European Union’s Horizon 2020 research and innovation programme and from Croatia, France, Germany, Greece, Italy, Netherlands, Portugal, Spain, Sweden, and Switzerland. The EPI-SGA2 project, PCI2022-132935 is also co-funded by MCIN/AEI /10.13039/501100011033 and by the UE NextGenerationEU/PRTR.
\end{acks}

\bibliographystyle{ACM-Reference-Format}
\bibliography{99-sdv}

\end{document}